# Leveraging generative artificial intelligence for simulation-based physics experiments: A new approach to virtual learning about the real world

Yossi Ben-Zion,[1,*] Turhan K. Carroll,[2,*,†] Colin G. West,[3,*] Jesse Wong,[3] and Noah D. Finkelstein[3]

[1]Department of Physics, Bar-Ilan University, Ramat Gan IL, 52900, Israel
[2]Department of Workforce Education and Instructional Technology, University of Georgia, Athens, Georgia 30602, USA
[3]Department of Physics, University of Colorado Boulder, Boulder, Colorado 80309, USA
*Co-first authors
†Contact author: tkcarroll@uga.edu

**ABSTRACT**. This study investigates the impact of a novel application of generative artificial intelligence (AI) in physics instruction: engaging students in prompting, refining, and validating AI-constructed simulations of physical phenomena. In a second-semester physics course for life science majors, we conducted a comparative study of three instructional approaches in a laboratory focused on electric potentials: (i) students using physical equipment, (ii) students using a prebuilt simulator, and (iii) students using AI to generate a simulation. We found significant group differences in performance on conceptual assessments of the laboratory content ($\eta^2 = 0.359$). Post-hoc analysis showed that students in both the AI-generated and prebuilt simulation conditions scored significantly higher on the conceptual assessments than students in the physical equipment condition. Students in these groups also reported more favorable perceptions of the learning experience. Finally, this preliminary study highlights opportunities for developing students' modeling skills through the processes of designing, refining, and validating AI-generated simulations.

# I. INTRODUCTION

Generative artificial intelligence (AI) is in the midst of transforming our educational practices. Already these tools appear to be able to solve most undergraduate physics problems [ref OpenAI Physics], succeed at many undergraduate conceptual assessments from introductory [ FCI/FMCE] to more advanced instruments [BEMA], and to support student learning and success – in some cases better than humans working using state-of-the-art curricula and pedagogical practices [1]. At the same time, there is increasing attention on where, when and how to use these tools within the physics teaching community [2, 3, 4, 5], and we have observed a wide variation in the approaches both faculty and students take in using generative AI. Given the apparent inevitability of the changes brought by AI tools, the physics education community ought to consider where these technologies are headed and how they can effectively be used to support new ways of learning in our course environments.

To such ends we present one research-validated approach to using generative AI productively in our physics classrooms. At heart, this approach draws from a longstanding notion of learning-by-teaching [5, 6]. In this case, the students are "teaching" generative AI, or more precisely prompting genAI to produce effective simulations for modeling physics phenomena. A cornerstone of the process is having students validate both the technical and scientific aspects of these simulations, and iteratively prompt (or "teach") the AI to produce increasingly accurate models of physics phenomena. Students can then use these simulations to further explore the physics content. Our hope is that this pedagogical approach is more-or-less evergreen, and applicable no matter what the capacities of genAI are or where it evolves to. Ultimately this approach is designed to support core learning objectives in undergraduate physics classes, but with the added benefits of teaching students how to use these new emerging technologies effectively and engaging them in authentic STEM practices related to model-building and evaluation.

This present work may be seen as an extension of early work with the PhET simulations [8], where we documented that working with simulations supported student learning about the physical world. In fact, students working first with simulators and then with real circuit components outperformed students who only worked with the real equipment. Students using the simulators did better both on measures of conceptual survey of related topics (series and parallel circuits) and on the ability to conduct an experiment – physically manipulate equipment, and describe the experiment and outcomes. The take-away here is not that simulations are more effective than laboratory equipment per se, but that exposure to simulations can be a valuable approach to prepare students for working effectively with laboratory equipment (which is a skill we still value). In this work, we similarly demonstrate a technique based in simulation usage–this time updated with the new capabilities of generative AI–which supports both conceptual understanding and desirable affective outcomes, while also offering the potential for developing scientific modeling skills.

The present study explores the possibility of using generative AI to support student engagement and understanding of basic physics phenomena. In particular, this paper examines:

1) How does the approach documented here (having students prompt, validate, and refine a simulation produced by generative AI) impact students' conceptual understanding as compared to using a prebuilt (AI designed) simulation or compared to using physical equipment in a traditional physics laboratory focused on equipotential lines?

2) How do students reflect on the value, ease of use, and enjoyment of using an AI-produced simulation around equipotential lines (again, as compared to prebuilt simulations and physical equipment)?

In this work, we further explore some preliminary discussion about the potential impact on modeling skills and other laboratory learning goals.

# II. METHODS

### A. Research design and topic selection

The study was conducted during the spring of 2025 at a large, R1 public university in the western United States. Data was collected in the second semester of an introductory physics for life sciences ("IPLS") course, taught without calculus. Initial enrollment for the course was 187 students, of whom 54.5% were non-male identifying. Students represented a mixture of different majors, primarily from life science disciplines and premed tracks; the largest of these groups was integrated physiology students (44%) followed by molecular, cellular, and developmental biology ("MCDB," 24%). Notably, despite being an "introductory" physics course, the student population contained no freshmen, consisting instead of sophomores (6%), juniors (28%), seniors (38%) and a significant "postbaccalaureate" cohort (28%) already possessing an undergraduate degree but returning to school to complete physics as a requirement for further postgraduate study such as med school, vet school, or dental school.

The intervention took place in the laboratory component of the course, which comprised two of the five credits of the course. Weekly labs were run by graduate student teaching assistants graded primarily for active participation and constituted 10% of the students' final course grade. The AI-enabled methods studied here were deployed during the third such laboratory meeting of the semester, with data collected immediately at the end of the lab period, on the subsequent midterm exam one week later, and on the final exam approximately two and a half months later. To examine differences in conceptual understanding and student perceptions resulting from different approaches to teaching the same physics content with varying forms of AI-engagement, a comparative study was designed with three quasi-experimental groups.

The study investigated the effects of:
- The unmodified "physical" laboratory which involved hands-on use of equipment to generate and measure real, physical quantities. This approach served as a control group.
- A lab involving use of a "pre-built" digital simulation, designed by author Ben-Zion using AI tools, but shared with the students only in its final form.
- An "AI-engagement" lab activity in which students used AI tools to generate and test their own simulation, but otherwise undertaking the same tasks as the students with the "pre-built" simulation.

The study created three independent groups, where each course participant was exposed to only one of the instructional approaches under investigation [9]. This design choice was made to ensure that comparisons reflect the unique influence of each approach without compromise from other course or student factors. There were seven laboratory sections in the course taught by four teaching assistants. Three teaching assistants were responsible for two sections each, with the fourth handling the remaining section. This fourth TA with only one section was assigned to teach the lab in the traditional, physical laboratory format; the remaining three were randomly assigned one section using the pre-built simulation and one section using the AI-engagement method.

The focal concepts for the labs during this period were electric potential and equipotential lines. The choice to deploy the interventions during this week was motivated by both the fundamental importance of electrostatic potential in the undergraduate physics curriculum and the specific pedagogical challenges it presents along with the opportunities of simulation and AI-assisted outcomes. Unlike more tangible physical concepts such as force or motion, electric potential is an abstract quantity which students may conceptualize in a variety of ways [10], presenting substantial pedagogical challenges. The topic is also often identified as one of the most important pathways to understanding subsequent topics in physics and engineering such as circuit analysis [11, 12], and failure to develop a familiarity with electric potential concepts can be a significant obstacle to understanding subsequent coursework [13].

### B. Activity objectives and description

The activities, in each of the three variations (physical, pre-built sim, and AI-engaged sim development), aimed to introduce students to the concept of electric potential through practical investigation of its spatial distribution around various charge configurations. The activity focused on three core learning objectives common to all groups.

The first objective was understanding the quantitative relationship between electric potential and distance from a point charge. Students were expected to examine the formula $V = kq/r$ through measurements at various points, compare measured values with theoretical results, and understand how potential varies with distance for both positive and negative charges.

The second objective focused on understanding equipotential lines and their properties for single charges and multiple charge configurations. Students were required to identify and map lines of constant potential, understand the relationship between line shapes and charge type and location, and investigate how multiple charges affect the resulting patterns. Additionally, they examined locations where potential becomes zero in systems with both positive and negative charges.

The third objective addressed understanding the relationship between equipotential line density and potential differences, as well as the physical significance of motion along these lines.

The physical laboratory group additionally investigated the effects of different charge geometries (non-point charges) on potential distribution and the explicit relationship between equipotential lines and electric field lines. Both simulation groups focused exclusively on point charges but included additional digital manipulation capabilities.

Across all groups, the activity concluded with an identical conceptual assessment consisting of five multiple-choice questions addressing the common concepts and learning objectives, and a feedback questionnaire regarding the student perspectives of the given medium of laboratory.

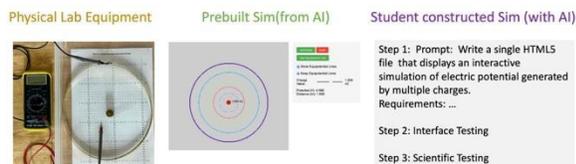

FIG 1. Three conditions of the laboratory: Physical Equipment, Prebuilt Simulation, AI Engaged Simulation Design

### Physical Equipment

Students in the traditional physical lab group took part in a "tried and true" activity which has been employed in essentially the same form as part of this coursework for a decade or more. The activity had been designed by members of the CU PER group, building on effective approaches in the physics community at the time. The equipment for the lab consisted of a large, shallow plastic storage tub filled with a mildly conductive solution. By placing metal objects at various points in the tub and connecting them to a battery to establish potential differences, students can create a variety of potential differences throughout the liquid, which can then be measured by a multimeter. Using graph paper positioned under the translucent base of the storage tub, they can then map out equipotential lines from various charge distributions to study their shape and properties.

For example, using a metal ring at the edges of the tub as "ground," and a narrow metal cylinder in the center as an approximate ``point charge,'' students can observe the classic 1/r concentric-circular equipotential pattern of a lone point charge on a 2D plane. In the activity, students first begin with this exploration, followed by further investigation of what happens under various changes to the system such as reversing the polarity of the battery or increasing the magnitude of the voltage difference between the central charge and the grounding ring. Subsequently, students were prompted to explore concepts relating potential difference to concepts like electric potential energy through the contrivance of an LED light with its leads connected to various points in the tub, identifying for example that if both leads fall on an equipotential, the bulb will not light up.

Finally, students explored more complex arrangements of the equipment to create scenarios equivalent to the presence of multiple point-charges, exploring the qualitative and quantitative effect on the equipotential lines, including identifying scenarios and specific points where the electric potential might vanish relative to ground.

### Prebuilt digital simulation

The pre-existing simulation provided students with an interactive digital interface for exploring electric potential. The tool enabled students to add point charges of both positive and negative sign to the workspace, adjust their magnitudes using a slider control, and reposition them by dragging. The simulation displayed real-time potential values and distances from charges as students moved the cursor across the screen. A key feature was the ability to generate and visualize equipotential lines by clicking on specific points, with options to display multiple lines simultaneously in different colors. The interface included reset and clear functions to remove charges or equipotential lines and restart the investigation.

The activity, while not running in identical sequence to the traditional lab activity, followed the same conceptual flow, and covered the same three activity objectives listed above, with modest variation in the nature and focus on each subtopic. The activity began with quantitative verification, where students placed a single charge and measured potential values at various distances, comparing these measurements with theoretical calculations using $V = kq/r$. This initial phase established familiarity with the simulation interface while reinforcing the mathematical relationship between potential and distance. Students then explored superposition effects by adding multiple charges and investigating locations where the net potential becomes zero, discovering that such points exist only when charges of opposite signs are present. The investigation proceeded to mapping equipotential lines, where students learned to visualize regions of constant potential by clicking on points and observing the resulting contours. Students systematically mapped multiple equipotential lines at regular voltage intervals for both positive and negative single charges, compared the resulting patterns, and subsequently investigated how the addition of a second positive charge altered the equipotential line geometry near each charge, between the charges, and far from both charges.

This approach emphasized guided exploration using a ready-made tool, allowing students to focus immediately on the physics concepts without technical barriers. The pre-existing simulation enabled rapid investigation of multiple scenarios and parameter variations, facilitating pattern recognition and conceptual understanding through iterative experimentation. Students could concentrate on interpreting results and making connections between mathematical relationships and visual representations, though they remained users rather than creators of the simulation environment.

**AI simulation design**

Students in the final group constructed their own electric potential simulation through structured interaction with Claude AI, running on its Sonnet 4 model, free version[1]. The activity required students to generate code through prompt engineering, systematically validate both interface functionality and physical accuracy, and iteratively refine the simulation through natural language feedback. This approach combined conceptual physics focus with computational framing, as students needed to articulate physical requirements precisely and verify that the resulting simulation adhered to established electrostatic principles.

The activity began with students receiving a comprehensive initial prompt designed to generate a complete electric potential simulation. This prompt served as the foundation for the entire learning experience, requiring students to copy and paste detailed specifications into Claude AI. The prompt read:

> Task: Write a single HTML5 file (including JavaScript and CSS) that displays an interactive simulation of electric potential generated by multiple charges.
>
> Requirements: Canvas Display: Create an element to display the simulation Use 2-pixel grid resolution for accurate equipotential lines
>
> User Interface: "Add Charge" button to place a new charge at the center of the canvas (+1 nanocoulomb default) Slider to adjust the selected charge value (-10 to +10 nanocoulombs, 1 steps) Display all numerical values with 3 decimal places (not in scientific notation) Display the current charge value dynamically next to the slider Allow dragging charges to reposition them "Reset" button to clear all charges
>
> Electric Potential: Calculate the potential at any point using the formula: $V = k * q / r$ k: Coulomb's constant ($9 \times 10^9$) q: Charge value in nanocoulombs (convert to SI units) r: Distance from the point to the charge Define minimum allowed distance from charges for potential calculation (to handle near-field behavior) Slider controls the value of the selected charge Use colors to distinguish positive (red) and negative (blue) charges Show each charge's value (e.g., "+1 nC") next to the charge When moving the cursor over the canvas, Display in the control panel the potential value (V) and in the case of 1 charge only: display also the distance from the charge (r) (in volts, meters)
>
> Charge Interaction: Click a charge to select it Slider controls the value of the selected charge Use colors to distinguish positive (red) and negative (blue) charges Show each charge's value (e.g., "+1 nC") next to the charge

Following code generation from this initial prompt, the activity evolved into a structured validation and enhancement process, while otherwise tracking the form of the activity with the prebuilt simulation. Students proceeded through systematic testing phases to verify both technical functionality and physical accuracy, followed by iterative refinement through natural language communication with the AI [2]. Notably, while the initial prompt producing the simulation was given to the students, all subsequent prompts were left open for the students to generate.

Following this design-and-verification phase, students were required to independently formulate additional prompts to extend the simulation's capabilities and address any issues that emerged during testing. This progression transformed the initial code output into a fully functional educational tool tailored to their specific learning objectives. Apart from the interaction with artificial intelligence, the creation and validation processes, the activities within the lab and physics content explored were identical to that of the pre-existing simulation.

**C. Assessment of impacts**

Students' understanding of the focal concepts of the laboratory was assessed at the end of the laboratory. Five conceptual questions ("CQs") were attached to the laboratory and completed by the students immediately after their lab activity. These questions are included in the Appendix 1.

---

[1] For those students who exhausted the number of allowed prompts (varying from 5 to 12) in the free version, Claude reverted to using Hiyku 3.5. Notably this caused some challenges for students, who ended up working with their peers or the slower, less powerful version of the AI engine.

Immediately after the lab, in addition to the conceptual questions described above, students were surveyed about their experiences in the laboratory and the particular approach taken (physical equipment, prebuilt simulation, AI-engaged development). Five Likert-scale questions probed student views:

1) How did you like this lab compared to last week's lab?
   Response: "1. way worse" to "5. way better"

2) How easy was it for you to use the [physical equipment/ simulation / AI]?
   Response: "1. extremely difficult" to "5. extremely easy"

3) Do you feel like the [physical equipment/AI/Sim] helped you understand Voltage / Electric Potential?
   Response: "1. Not at all" to "5. Enormously"

4) Did you enjoy working with the [physical equipment/ AI/Simulator]?
   Response: "1. Really didn't like" to "5. A great deal"

5) Would you suggest we do this again?
   Response: "1. Definitely no" to "5. Definitely yes"

Students were also given space to write either explanations for their Likert-scale responses or other unprompted sentiments.

Student performance on the midterm, three weeks later, and the final, 12 weeks later, was also captured. Data were collected both on overall performance on the midterm (MQs) and final (FQs). These data were designed to document any overall differences in the samples. That is, given the common homework, interactive lectures, and study sessions provided to all students after the laboratory experience, we expected no difference on student performance on the midterm and final. These measures were used to document similarity of samples.

### III. DATA ANALYSIS AND RESULTS
#### A. Analysis of student performance on the lab review, midterm exam, and final exam

We performed several statistical tests to understand the relationship between lab type and performance on the three sets of content questions described above (CQs, MQs, and FQs). Since this was an exploratory study with the goal of comparing more than two groups along a categorical variable (the type of lab experience, "Lab Type"), we performed an omnibus test to assess whether there were any statistically significant differences between groups in the outcomes of interest. Table I shows the variables we used for our omnibus tests.

Given that research question 1 aims to compare assessment performance across lab types, we planned to use one-way ANOVA to assess mean performance differences across lab types. We performed preliminary analysis to see whether our data satisfied the 6 assumptions of ANOVA. The details of this preliminary analysis are provided in Appendix 2. We concluded that, while most of the assumptions are satisfied, the residuals of the dependent variables are not normally distributed. As a result, we utilized nonparametric statistical techniques in our analysis. We compared median differences across lab type as median is the appropriate measure of central tendency for nonparametric data [14].

In order to visually represent the medians $CQ_{tot}$, $MQ_{tot}$, and $FQ_{tot}$ across lab groups and highlight potential median differences, we created bar charts (with error bars representing the standard error of the median). These bar charts are shown in Figures 2-4 below. Standard errors of the medians were calculated using bootstrapping with replacement [15, 16]. Our procedure used 1000 bootstrap replicates. This method for calculating confidence intervals was used because of the non-normality of our data.

TABLE I. Variables used for omnibus tests.

| Variable | Level of Measurement | Definition | Value Range |
|---|---|---|---|
| Lab Type | Nominal | The type of lab a student participated in. | AI (student generated AI sim); Sim (pre-built AI sim); Lab (traditional lab) |
| $CQ_{Tot}$ | Interval | Total score on the review questions students completed about electric potential as part of their lab activity. | 0-100 |
| $MQ_{Tot}$ | Interval | Student's total score on the midterm exam. | 0-100 |
| $FQ_{Tot}$ | Interval | Student's total score on the final exam. | 0-100 |

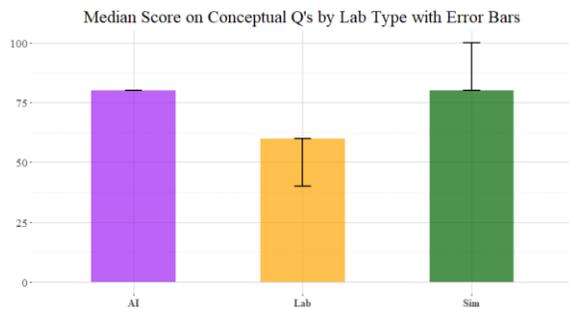

FIG 2: Median scores for the conceptual questions ($CQ_{tot}$). Error bars represent the standard error of the median.

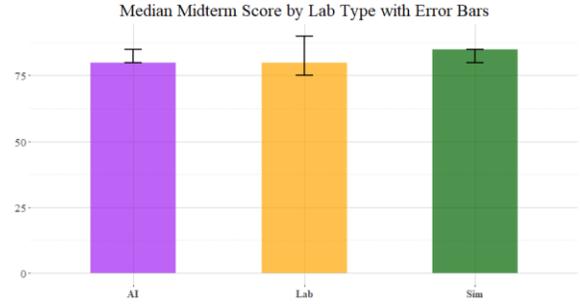

FIG 3: Median scores for the midterm exam ($MQ_{tot}$). Error bars represent the standard error of the median.

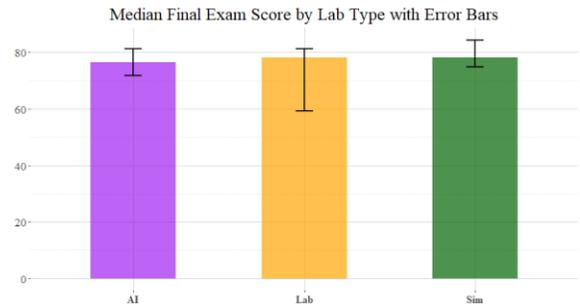

FIG 4: Median scores for the final exam ($FQ_{tot}$). Error bars represent the standard error of the median.

The figures above suggest that there are large differences in medians of $CQ_{tot}$ between the physical lab and other conditions, and small differences in the medians for $MT_{tot}$ and $FE_{tot}$. In order to quantify the significance of these differences, we used Kruska-Wallis test, a nonparametric omnibus test which determines whether there are statistically significant differences between the medians of three or more groups and does not assume normality of residuals [17].

The results of our Kruskal-Wallis analyses are in Table II below:

TABLE II. Kruskal-Wallis Test Results for each Dependent Variable.

|  | df | H-Statistic | p-Value | $\eta^2$ |
|---|---|---|---|---|
| $CQ_{Tot}$ | 2 | 58.718 | 0.000 | 0.359 |
| $MQ_{Tot}$ | 2 | 0.837 | 0.658 | N/A |
| $FQ_{Tot}$ | 2 | 3.302 | 0.192 | N/A |

We found that median performance on CQs differed significantly across our three lab types, $H_{CQTot}(2)=58.718, p<.001$. We found that there was no significant difference in median MQ performance across lab type, $H_{MQTot}(2) = 0.837, p = .658$, and the differences in FQ scores across lab types were not found to be significant at the .05 significance-level ($H_{FQTot}(2)=3.302, p=.192$).

The analysis in Table II establishes that the median CQ scores–that is, the conceptual questions answered by students immediately after their labs–showed significant differences across the three groups. We used $\eta^2$ to assess the effect size of these differences [18] and found that the effect was strong ($\eta^2=.359$). In order to assess which groups showed significant differences in performance, we used Dunn's post-hoc test [19]:

TABLE III. Results of Dunn's Test for $CQ_{Tot}$.

| Comparison | Adjusted p-value |
|---|---|
| AI vs Lab | 0.000 |
| AI vs Sim | 0.258 |
| Sim vs Lab | 0.000 |

We can see from Table III that there are no significant group differences in performance on CQs between students in the two simulation lab sections (AI and prebuilt sim). However, there are statistically significant differences between the AI and Lab students ($p_{AI\text{-}lab\ Adj} < .001$) and between Sim and Lab students ($p_{Sim\text{-}lab\ Adj} < .001$). We can conclude that students in the AI and Sim lab groups both performed significantly better than students in the traditional lab section.

Since this was an exploratory study, we also performed a question-by-question analysis of the conceptual questions to see if the results were consistent with our omnibus analysis. The results of this analysis, which is included in Appendix 3, are consistent with the results reported above.

### B. Analysis of Student Attitudes Toward Laboratory Condition

We then examined responses to ascertain student attitudes and perspectives on the laboratories (AQs) we included at the end of their lab activity. These questions were analyzed individually as they were not designed to form a singular construct. There were 5 questions, and each was meant to probe students' attitudes after completing their lab activity. As appropriate, the wording used for the questions was varied to address the particular lab setting in which the student worked. The questions used a 5-point Likert scale (strongly disagree, disagree, neutral, agree, strongly agree), and each question was coded in a way that higher score indicated more positive affect. For our analysis, we collapsed the strongly disagree/disagree categories into one category and we collapsed the strongly agree/agree categories into one category because upon discussion, the research team believed the varying "agree" and "disagree" categories to be redundant. Prior work has suggested collapsing the five-point ordinal scale to a three-point ordinal scale, can be done in cases where respondents may use the various "agree" and "disagree" categories redundantly [20]. Table IV lists the variables used for this analysis:

TABLE IV. Variables Used for our Question-by-Question Analysis of the Student Attitude Questions.

| Variable | Level of Measurement | Definition | Value Range |
|---|---|---|---|
| Lab Type | Nominal | The type of lab a student participated in. | AI; Sim; Lab |
| $AQ_n$ | Ordinal | Review question n, where n=1,2,3,4, or 5. | 1 = Disagree; 2 = Neutral; 3 = Agree |

This analysis was done in three stages. First, contingency tables were developed depicting the frequency of each response option for each AQ within each lab type to inform us about how many students in each lab type selected disagree/neutral/agree. This analysis of each question resulted in a contingency table, for example, the contingency table for $AQ_1$ is shown in Table V below (other contingency tables are available upon request).

TABLE V: Contingency Table for $AQ_1$

| | Agree | Disagree | Neutral |
|---|---|---|---|
| AI | 48.00 | 4.00 | 14.00 |
| Lab | 5.00 | 6.00 | 16.00 |
| Sim | 45.00 | 0.00 | 23.00 |

For the second phase of the analysis, we statistically tested the association between lab type and student response. This was done using Fisher's exact test because our contingency tables had cells containing a frequency that was less than 5 [21]. Cramer's V [18] was used as a measure of effect size for this test. Our results for each AQ are listed in Table VI below.

TABLE VI: Fisher's Test Results for Each AQ.

| Question | p-Value | Cramer's V |
|---|---|---|
| $AQ_1$ | 3.581e-7 | 0.323 |
| $AQ_2$ | 7.257e-14 | 0.475 |
| $AQ_3$ | 0.005 | 0.215 |
| $AQ_4$ | 0.002 | 0.25 |
| $AQ_5$ | 8.969e-7 | 0.323 |

As the p-value (two-tailed) obtained from Fisher's exact test is significant for each AQ (see p-values and effect sizes in Table VI), we see that there is a statistically significant association between lab type and student response for each of our student attitude questions.

Though the Fisher's test established a significant association between lab type and student attitudes, it does not reveal which lab types have a strong association. Given that each of our contingency tables in this analysis were 3x3 (three lab types and three possible outcomes), in the third phase of our analysis we performed a post-hoc pairwise Fisher's exact test to compare each lab type. The results of the pairwise comparisons are tabulated in Table VII:

TABLE VII: Pairwise Comparisons for Affective Questions.

| | $p_{adjAQ1}$ | $p_{adjAQ2}$ | $p_{adjAQ3}$ | $p_{adjAQ4}$ | $p_{adjAQ5}$ |
|---|---|---|---|---|---|
| AI vs Sim | **0.035*** | 0.245 | 0.343 | 0.688 | 0.658 |
| AI vs Lab | 9.23e-6*** | 4.8e-10*** | 0.037* | 0.002** | 1.38e-5*** |
| Sim vs Lab | 3.57e-6*** | 2.81e-13*** | 0.002** | 0.002** | 9.66e-7*** |

* p - .05; ** p - .001; *** p<.001
Pairwise p-values indicating whether or not there was a significant difference between the indicated pair of treatment groups (rows) on a particular attitude question (columns). Purple = AI performed better; Orange = Lab performed better; Green = Sim performed better; Black = No significant difference.

Fisher's pairwise exact tests indicate that students who did the AI lab reported more positive attitudes than students who did the traditional lab for all AQs ($p_{adjAQ1,AI\text{-}Lab} < .001$, $p_{adjAQ2,AI\text{-}Lab} < .001$, $p_{adjAQ3,AI\text{-}Lab} = .037$, $p_{adjAQ4,AI\text{-}Lab} = .002$, $p_{adjAQ5,AI\text{-}Lab} < .001$). Similarly, students who did the lab using pre-built simulations also reported more positive attitudes than students who did the traditional lab ($p_{adjAQ1,Sim\text{-}Lab} < .001$, $p_{adjAQ2, Sim\text{-}Lab} < .001$, $p_{adjAQ3, Sim\text{-}Lab} = .002$, $p_{adjAQ4, Sim\text{-}Lab} = .002$, $p_{adjAQ5, Sim\text{-}Lab} < .001$). Comparing between the two simulation groups, however (AI and prebuit), there is significant difference only for $AQ_1$ ($p_{adjAQ1,AI\text{-}Sim} = .035$). We can conclude that students who did the lab via pre-built simulations demonstrated more positive self-reported attitudes on $AQ_1$ than students who did the lab using AI.

## IV. DISCUSSION
### A. Developing conceptual understanding

The interactive use of a generative AI platform, where students prompt, validate and refine a simulation, can support conceptual understanding of equipotential lines as effectively as a pre-built simulation and significantly better than using physical equipment. From the results we see a significant and sizable impact on conceptual understanding based on the approach taken (p<0.001, $\eta^2$ = .359) among the three experimental conditions. In pairwise comparisons, we observe significant differences between the physical equipment group and each of the other two approaches, pre-existing sim and AI-engaged laboratory. There was no significant difference between the pre-existing sim group and the AI-engaged group. Furthermore, in a similar analysis (not reported here), we found no significant differences in student midterm performance, or final exam performance across lab sections or lab instructors (indicating a solid basis for comparison). Hence, there is strong indication that both the use of the prebuilt simulation, and the engaged use of generative AI to develop, refine and validate a simulation, were comparably productive in developing student understanding of concepts related to the laboratory.

In one sense, these results replicate the main results of earlier studies where simulations can support greater conceptual understanding of physics concepts in the immediate aftermath of a learning activity [7]. In these earlier studies, the learning differences were also found to persist through the end of the term, which we notably did not replicate here. As observed above, the student groups did not perform differently on the midterm / final overall; in fact, there was also no variation in performance between groups even considering only the midterm and final questions focused on electric potential as a concept. That is, the differences in conceptual performance difference among the treatment groups that showed up following the lab vanished on the subsequent midterm and final exam questions. However, a significant difference in the broader course context between this and the 2005 study offers a plausible explanation of this difference: in the prior studies, the simulation lab which was studied took place only after the relevant topic had been fully discussed in lecture, and was part of a course which did not employ modern interactive teaching methods. In this current work, the lab we studied was followed by continued discussion of the content in lectures, homework, and exam reviews, including interactive and peer-instructional methods associated with improved learning outcomes [22, 23, 24].

And yet, beyond the confirmation of the prior results for the simulation groups, our comparable results for the AI-engaged groups is striking. It is far from given that students' use of generative AI to build simulations would support their conceptual development. For one thing, building the simulation with the aid of the AI is an additional layer of work on top of the actual use of the simulation itself. For another, AI is prone to hallucination, or presenting material not-well matched to undergraduate learners [25]. Nor are these learners particularly well-prepared in simulation development or use of generative AI, and we did not provide any instruction on these topics prior to the lab activities documented here. However, we do find that student prompted, refined and validated simulation development did promote conceptual understanding just as well as when they used a prebuilt, vetted and validated simulation, and to a greater extent than when they used real equipment. In short, it appears that at least under appropriate conditions, the opportunity to work with AI tools can be added without cost to conceptual learning.

### B. Student attitudes

Paralleling what was observed in the CQs, AQ responses indicated that students appreciate both the pre-built simulation and their own sim with an AI more than using the physical equipment. On each of the AQs, students using the digital technologies reported more favorable responses than those using the physical lab equipment. Of course, using the pre-built simulation was faster and easier than either of the other two conditions. So, some students noted this affordance and appreciated being done with the laboratory sooner. This may also account for the one instance where students reported preference for the use of the prebuilt sim over the use of the AI generation of a sim, on AQ1 comparing this approach to prior weeks.

In other cases, students reflected that the prebuilt simulation and the design-based approach supported their conceptual learning; whereas the equipment was less useful to such ends. A student noted, "One of the reasons I like the simulation compared to a hands-on experiment is that it tends to be more difficult for me to carry out and understand a hands-on experiment, whereas with the simulation I feel like more time is spent thinking through and understanding the material. In this sense, we agree with them. Working with physical equipment may be better used to facilitate experimental and modeling skills rather than at reinforcing theory or concepts.

Finally, both the group building an AI-based simulation and the group using physical equipment sometimes reflected frustration in "getting it right" or "fixing the equipment / AI". While this can be problematic if students are overly distracted or disengage as a result, it can also be a positive side capturing the productive frustration that learning can entail. While students expressed frustrations in developing their AI simulations, students' frustrations were similar to some of the frustrations of students working with physical equipment– around design and manipulation of the materials. None of the student concerns focused on a limited opportunity to learn or engage with physics concepts. This sentiment is in contrast to the group using physical equipment, which expressed similar frustrations with the equipment with clear indication that they believed their learning was negatively impacted. Such sentiments also highlight the capacities for an AI environment to develop modeling and debugging skills that we seek to foster with less negative impact on the students' conceptual understanding of the material.

### C. An opportunity for modeling and experimental skills

Our study has focused on conceptual learning, which, to many, is a valuable learning goal in the kind of lab- or recitation-like settings which might make use of simulations as an instructional tool. However, another perspective suggests the focal learning outcomes in such settings could be the development of scientific modeling practices, regarded by some as an even more natural objective for laboratory instruction [26, 27, 28]. While beyond the scope of this paper, we suspect that the approach taken to prompting, refining, and validating AI-generated simulations can develop relevant scientific and modeling skills that we do seek to promote in our laboratory environment.

In this case, rather than considering the traditional laboratory with students using physical equipment as a control condition, here we might consider the use of the pre-built simulation as a control condition, as both the equipment group and the simulation creation group have the opportunity to develop experimental skills. We are not arguing that the *same* skills are practiced in each group, though they may be analogous. For example, students interacting with the pre-built simulation and students interacting with the AI-generated simulation both reported (in free-response feedback) an appreciation for the ability to visualize the intangible or unseen such as equipotential lines. However, the students using the pre-built simulation did not have the opportunity to validate (technically or scientifically) the tool they were using.  This validation may parallel the modeling skills which [27] Lewandowski et al and others call for -- technical validation as analogous to modeling of physical apparatus, and scientific validation as analogous to the physical modeling.  By interacting with the AI to build these representations, students building a sim with AI had more opportunities to learn skills similar to the experiment group by developing and applying models and validating them to aid their understanding.

Though the data presented in this work are not intended to speak directly to the question, we speculate about the mechanism by which that the iterative, validation phase of AI-based lab work, which was not present for students working with the prebuilt simulation could contribute to improved modelling skills in the AI-engaged groups. This phase specifically invited students to view the simulations generated from their prompts with skepticism. In this case the fact that AI is based on a probabilistic engine (and even that it is subject to hallucinations) can serve as an affordance.  By constructing imperfect results, or perhaps just the possibility of producing imperfect results, has the natural affordance that requires (or encourages) students to engage in validation and verification in ways that other mediums may not.  In the pre-built simulation groups, students were essentially asked to treat the simulations as directly reflecting the true physical principles governing the underlying phenomena. In doing so, the nature of the simulation as a "model" is obscured, since the students do not necessarily consider any distinction between the simulation and the physical reality it is meant to represent. A model by its nature should have predictive capability but also inherent limitations in its sphere of applicability [29]; when students use pre-existing simulations in coursework, they may be a danger of overlooking these aspects of the simulation-as-model if the simulation accuracy is simply taken for granted By contrast, students in the AI-engaged lab activity were asked to treat each iteration of simulation output by the AI as a possible model of the real world, but one whose fidelity remained to be established even in the simplest contexts before it could be responsibly used to explore more complicated scenarios.

### D. Limitations

Several methodological and contextual factors constrain the interpretation and generalizability of these findings. First, while all three instructional approaches addressed electric potential concepts through active engagement, systematic differences existed in the level of abstraction required. In the physical laboratory approach, for example, the "basic" objects presented to the students for use and consideration were the source charges and the conductive bath around them in which potential differences could be measured and from there, equipotential shapes inferred. In the simulations, the equipotentials themselves were presented to students as part of the "basic" objects available for their contemplation–available instantly at the push of a button, just like the source charges which create them. Although each approach incorporated both conceptual reasoning and quantitative analysis, this fundamental difference in abstraction level of abstraction may have systematically influenced performance on the conceptual assessment instrument used in this study, which may be considered a limitation on the one hand, or a finding on the other.

Second, technical limitations constrained the AI-engaged group's implementation fidelity. Students encountered restrictions imposed by the free-tier service limits of the AI platform, including constraints on prompt frequency and session duration. Additionally, some participants required instructional assistance to resolve technical issues during simulation development or equipment use, potentially affecting the consistency of the intervention delivery across participants.

Third, this investigation examined a single physics concept (electric potential) within a homogeneous student population at a single research institution. Participants comprised primarily life sciences majors with similar academic backgrounds and career trajectories. The extent to which these findings generalize to other physics topics, more diverse student populations, or different institutional contexts remains an empirical question requiring systematic investigation across varied educational settings.

### V. CONCLUSIONS

The use of generative AI to design, refine and validate a simulation can support student conceptual mastery of key physics topics in a laboratory environment. Students who developed simulations through AI-guided processes achieved conceptual understanding equivalent to those using pre-built simulations, with both digital approaches significantly outperforming traditional physical laboratory methods. Notably, despite the additional cognitive demands of prompting, validating, and refining AI-generated simulations, students' conceptual learning was not compromised.

These preliminary findings indicate three complementary pathways for physics laboratory education, which by its nature pursues a wide variety of learning outcomes beyond the basic development of theoretical knowledge. Pre-built simulations effectively engage students in learning key concepts while providing immediate access to complex visualizations. AI-guided simulation development supports conceptual learning, an awareness of emerging technologies, and an opportunity to model systems. Traditional physical equipment remains essential for developing experimental skills with physical equipment, modeling, and connecting students to the material world that physics seeks to understand.

Given that generative artificial intelligence will likely constitute a significant component of students' future professional and academic lives, developing competencies in productive AI use represents an important educational objective. Our results demonstrate that this can be achieved without compromising traditional physics learning goals.

Physics fundamentally concerns the study of the material world, and experimental equipment use remains essential for developing authentic scientific practices. Generative AI now provides a complementary tool for use alongside physical equipment and pre-built simulations. The chosen mixture of among these approaches should align with context-specific learning objectives, recognizing that different methods foster distinct forms of active learning and skill development.

Future investigations should examine the pedagogical framing required for optimal implementation of these tools. Neither excessive optimism nor cynicism is warranted; these tools require careful implementation with attention to their limitations, but can demonstrably serve student learning and engagement when thoughtfully deployed.

As physics education evolves to meet the demands of an increasingly digital world, the integration of AI-assisted learning alongside traditional experimental work offers a promising framework. This complementary approach may help prepare students with the diverse competencies needed for contemporary scientific practice.

# APPENDIX

## APPENDIX 1: Survey of conceptual understanding

Circle your answers to the following questions, to check your understanding). In all questions, use the convention that the potential is zero infinitely far from the charges.

1. A large positive charge sits alone in the center of my lab (no other charges are present). At a certain point "P" which is two meters away from the positive charge, I measure the potential to be 40V. What is the potential at point "Q" which is one meter away from the positive charge?

A. 10V
B. 20V
C. 40V
D. 80V
E. 160V

2. Four different (nonzero) electric charges are placed in a square. At the exact center of the square, the potential is exactly zero. What can we say about the signs of the charges? Select one.

A. All of the charges must be positive
B. At least one of the charges must be negative
C. Exactly two of the changes must be negative
D. All of the charges must be negative
E. None of the above is necessarily true

3. In which of the following cases would the equipotential lines around the charges form perfect circles? Assume there are no other charges present besides the ones described. Select one

A. A single positive charge is located at the origin
B. A single negative charge is located at the origin
C. A single positive charge is located at (x = 3m, y = 3m)
D. Both (A) and (B) but not (C)
E. (A), (B), and also (C)

4) What happens to the shape of equipotential lines when switching the sign of every charge (positive ⇌ negative) in a system?

A. The equipotential lines flip and completely change their shape.
B. The lines remain the exact same shape, but their values switch from positive to negative (or vice versa).
C. The equipotential lines for a negative charge are more closely spaced than those for a positive charge.
D. There is no change in equipotential lines.
E. We can't tell what happens
F. None of the above

5. Consider the two positive charges depicted at right, along with the equipotential line shown around them. What must be true about the magnitude of the charges?

A. Q1 = Q2
B. Q1 > Q2
C. Q1 < Q2
D. Cannot determine without more information

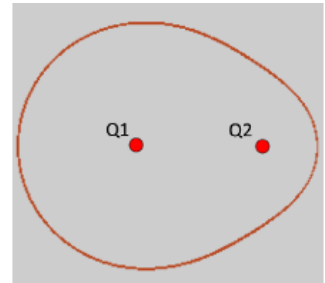

**APPENDIX 2: Testing ANOVA Assumptions**

The assumptions for a one-way ANOVA test are:

1. Our dependent variable is continuous: this is true for our variables.
2. Our independent variable consists of 2 or more categorical, independent groups: this is true for our data.
3. Independence of observations: This is true to the best of our knowledge.
4. No significant outliers in the data:
5. The residuals of each dependent variable are approximately normally distributed for each category.
6. Homogeneity of variances across categories.

Assumptions 1-3 are matters of design. Assumptions 1 and 2 follow directly from our study design. Assumption 3 is approximately true to the best of our knowledge. Graphical inspection of our data showed that there were 8 outliers in our dataset across all groups. This represents ~5% of our data and it was determined that these outliers did not cause a significant distortion of our results, thus assumption 4 held for our data. In order to test assumption 5, we calculated the skewness and kurtosis for our dependent variables across each category. These results are documented in Table A2.1:

TABLE A2.1: Descriptive statistics for each lab type

| Lab Type | | $CQ_{Tot}$ | $MQ_{Tot}$ | $FQ_{Tot}$ |
|---|---|---|---|---|
| AI (N=66) | Median | 80 | 80 | 76.6 |
| | Mean | 83.64 | 81.82 | 74.86 |
| | Std. Deviation | 14.43 | 12.998 | 14.10 |
| | Skewness | -0.543 | -0.441 | -0.420 |
| | Kurtosis | 0.008 | -0.198 | -0.573 |
| Sim (N=68) | Median | 80 | 85 | 78.1 |
| | Mean | 85.59 | 81.69 | 77.11 |
| | Std. Deviation | 13.76 | 13.43 | 14.63 |
| | Skewness | -0.428 | -0.883 | -1.390 |
| | Kurtosis | -0.815 | 0.753 | 3.492 |
| Lab (N=27) | Median | 60 | 80 | 78.1 |
| | Mean | 48.89 | 78.15 | 68.98 |
| | Std. Deviation | 20.25 | 17.05 | 19.42 |
| | Skewness | -1.517 | -0.430 | -0.585 |
| | Kurtosis | 1.907 | -0.786 | -1.106 |

The skewness and kurtosis for each dependent variable is between -2 and 2 with one exception (the kurtosis for the FEScore for the Sim (pre-built simulation) group). Upon seeing this inconsistency, we statistically tested for normality for each dependent variable using the Shapiro-Wilk (SW) test [30, 31]. For the SW test, the null hypothesis is that a sample is drawn from a normally distributed population, so if $p < .05$ we conclude that our sample does not satisfy the assumption of normality. The SW test indicated that none of these variables follow a normal distribution for our sample ($W_{CQ} = .87$, $p_{SW-CQ} < .001$; $W_{MQ} = .96$, $p_{SW-MQ} < .001$; $W_{FQ} = .95$, $p_{SW-FQ} < .001$). In order to test for homogeneity of variances (assumption 6) we performed Levene's test of homogeneity of variances for our dependent variables. The results are shown in Table A2.2:

TABLE A2.2: Levene's Tests Results

|        | F-Value | $df_1$ | $df_2$ | p-Value |
|--------|---------|--------|--------|---------|
| $CQ_{Tot}$ | 0.335   | 2      | 158    | 0.717   |
| $MQ_{Tot}$ | 1.686   | 2      | 158    | 0.189   |
| $FQ_{Tot}$ | 2.45    | 2      | 158    | 0.09    |

$CQ_{Tot}$, $MQ_{Tot}$, and $FQ_{Tot}$ all show homogeneity of variances according to the Levene's test, ($p_{Lev-CQTot}(2,158)=.117$, $p_{Lev-MQTot}(2,158)=.169$, and $p_{Lev-FQTot}(2,158)=.008$).

# APPENDIX 3: Question-by-question analysis of CQs

*We performed analysis of both the CQs and the affective questions on a question-by-question basis using a three-phase method outlined below. In this appendix, we demonstrate the technique in the context of the CQs; equivalent data for the affective questions is available on request.*

Since our analysis of the $CQ_{Tot}$ indicated significant differences in performance, among the three lab types, on the lab review, we decided to analyze each review question individually to examine whether there were differences in the number of correct responses across each lab type. The variables for this analysis listed in Table A3.1:

TABLE A3.1: Variables used for question-by-question analysis of conceptual questions.

| Variable | Level of Measurement | Definition | Value Range |
|---|---|---|---|
| Lab Type | Nominal | The type of lab a student participated in. | AI; Sim; Lab |
| $CQ_n$ | Nominal | Conceptual question n, where n=1,2,3,4, or 5. | 0 = Incorrect; 1 = Correct |

For each question there were cells in our contingency table that had a frequency of less than 5, so in order to perform such an analysis. We used the same three phase analysis that was used for the question-by-question analysis of the student attitude questions detailed in the Data Analysis and Results section above. This analysis was done in three stages. First, contingency tables were developed depicting the frequency of correct and incorrect responses for each CQ within each lab type to inform us about how many students in each lab type got the question correct/incorrect. For example, the contingency table for $CQ_1$ is shown in Table A3.2 (other contingency tables are available upon request).

TABLE A3.2: Contingency Table for $CQ_1$

|  | Correct | Incorrect |
|---|---|---|
| AI | 63.00 | 3.00 |
| Lab | 5.00 | 22.00 |
| Sim | 66.00 | 2.00 |

The results of the Fisher's test for each CQ is in Table A3.3:

TABLE A3.3: Fisher's Test Results for Each CQ

| Question | p-Value | Cramer's V |
|---|---|---|
| $CQ_1$ | 2.2e-16 | 0.778 |
| $CQ_2$ | 0.002 | 0.268 |
| $CQ_3$ | 0.037 | 0.2 |
| $CQ_4$ | 0.009 | 0.22 |
| $CQ_5$ | 3.7e-7 | 0.481 |

As the p-value (two-tailed) obtained from Fisher's exact test is significant for each CQ (see p-values and effect sizes in Table A3.3), we reject the null hypothesis ($p < 0.05$) and conclude that there is a statistically significant association between lab type and student performance for each of our review questions.

The results of the pairwise comparisons are tabulated in Table A3.4:

TABLE A3.4: Pairwise Comparisons for Conceptual Questions

| | $p_{adjCQ1}$ | $p_{adjCQ2}$ | $p_{adjCQ3}$ | $p_{adjCQ4}$ | $p_{adjCQ5}$ |
|---|---|---|---|---|---|
| AI vs Sim | 0.68 | 0.86 | 0.37 | 0.025* | 1.00 |
| AI vs Lab | 1.86e-13*** | 0.003** | 0.07 | 1.00 | 4.65e-6*** |
| Sim vs Lab | 2.84e-13*** | 0.003** | 0.33 | 0.033* | 5.02e-6*** |

\* p - .05; \*\* p - .001; \*\*\* p<.001
Purple = AI performed better; Orange = Lab performed better; Green = Sim performed better; Black = No significant difference

Results of Fisher's pairwise exact tests indicate that there is significant association between the AI and Sim lab groups for $CQ_4$ ($p_{adjCQ4,AI-Sim} = .025$). We can conclude that students who did the lab via pre-built simulations were more likely to answer $CQ_4$ correctly than students who did the lab using AI. Students who did the AI lab were more likely to answer $CQ_1$, $CQ_2$, and $CQ_5$ correctly than students who did the traditional lab ($p_{adjCQ1,AI-Lab} < .001$, $p_{adjCQ2,AI-Lab} = .003$, $p_{adjCQ5,AI-Lab} = .005$). Students who did the lab using pre-built simulations were more likely to answer $CQ_1$, $CQ_2$, $CQ_4$, and $CQ_5$ correctly than students who did the traditional lab ($p_{adjCQ1,Sim-Lab} < .001$, $p_{adjCQ2, Sim-Lab} = .003$, $p_{adjCQ4, Sim-Lab} = .033$, $p_{adjCQ5, Sim-Lab} < .001$)